\documentstyle[11pt]{article}
\begin{document}
\title{On free differentials on associative algebras \footnote{Talk
presented by V. Kharchenko at the Third International Conference
in Non Associative Algebras and Its Applications, Oviedo, Spain, July 1993}}
\author{
A. Borowiec \thanks{Supported by the Polish Committee for Scientific Research
No 2~ 2419~ 92~ 03}\\	Institute of Theoretical Physics \\
University of Wroc{\l}aw, Poland, borowiec@plwruw11.bitnet
\and
V. K. Kharchenko \thanks{Supported by the Russian Fund of Fundamental
Research No 93-011-16171}\\ Institute of Mathematics\\
Novosibirsk, Russia, kharchen@math.nsk.su
\and
Z. Oziewicz \thanks{On leave from the Institute of Theoretical Physics
University of Wroc{\l}aw}\\
Universidad Nacional Autonoma de M\'exico\\ M\'exico ,
oziewicz@redvax1.dgsca.unam.mx}
\date{\bf November 93, ITP UWr 861/93}
\maketitle
\begin{abstract}
A free differential for an arbitrary associative algebra is
defined as a differential with a uniqueness property.
The existence problem for such a differential is posed. The notion
of optimal calculi for  given commutation rules is introduced
and an explicit construction of it for a homogenous case
is provided. Some examples are presented.
\end{abstract}

\def\c{\cdot}
\def\ot{\otimes}
\def\rt{\rightarrow}
\def\ld{\ldots}

\def\ba{\begin{array}}
\def\ea{\end{array}}

\def\R{\hat{R}}
\def\I{\hat{I}}
\def\x{\hat{x}}
\def\A{\hat{A}}
\def\pih{\hat{\pi}}
\def\D{\hat{D}}
\def\d{\hat{d}}
\def\al{\alpha}
\def\la{\lambda}
\def\ga{\gamma}

\section{Introduction}
A differential $d:R \rt\, _{R}\!M\!_{R}$ is called free if the differential
of any element $v$ has a unique presentation of the form $dv=dx^{i}\c v_{i}$,
where $x^1,\ld ,x^n$ are generators of the algebra and $dx^1,\ld ,dx^n$ their
differentials. Any free differential defines a commutation formula
$vdx^i=dx^k\c A(v)^i_k$, where $A:v \mapsto A(v)^i_k$ is an algebra
homomorphism $A:R \rt R_{n \times n}$. It is easy to see that for any
homomorphism $R \rt R_{n \times n}$ there exists not more than one
free differential. We are going to consider the existence problem of such a
differential. We will show that for a given commutation rule
$vdx^i=dx^k\c A(v)^i_k$ a free algebra generated by the variables $x^1,\ld
,x^n$
has a related free differential. We will define an {\it optimal} algebra with
respect to a fixed commutation rule. In the homogeneous case this algebra
is characterized as the unique algebra which has no nonzero $A$-invariant
subspaces with zero differentials. Finally, we will consider a number of
example
of optimal algebras for different commutation rules. In particular, we will
desc
two variable commutation rules which define commutative optimal algebra.

This article is closely related to the known Wess and Zumino paper
{\cite{Wess}. In our terms, they prove in particular that a system on $n^2$
quadratic forms vanish in the optimal algebra if the Yang-Baxter equation
holds.
\section{Free differential calculi}
Recall that a differential is a linear mapping from an algebra $R$
to a bimodule $M$ satisfying the Leibniz rule:
$$
d(uv)=d(u)v + ud(v)
\medskip  $$
{\bf Lemma 1.1} \em
A differential $d$ has the uniqueness property iff  $\Omega_d(R)=R^{\flat}
d(R)R^{\flat}$ is a free right $R$-module freely generated  by $dx^1,\ld ,
dx^n$  (here
$R^{\flat}=R$ if $R$ has a unit and the augmented algebra otherwise).
\medskip \\ \em
Due to the lemma the following definition is natural. \smallskip \\
{\bf Definition 1.2}
A differential is said to be {\it free} if it has the uniqueness property.
\medskip \\
That definition essentially depends on the generating space $V=\sum x^i F$.
Let us consider as example the case $V=R$. Of course, if $R$ is not finite
dimensional, then we have an infinite set of generators. Nevertheless, there
exists a free differential with respect to that space of generators. It is
exact
the universal derivation.

If $d$ is a free differential, then linear maps $D_k:R\rt R$ (partial
derivatives) can be defined by the formula:
$$
d\,v=dx^k\c D_k(v)
\eqno		   (1)
$$
Those maps satisfy the relations
$$
D_k(x^i)= \delta^i_k ,
\eqno	  (2)
$$
where	$\delta^i_k$ is the Kronecker delta.\medskip \\
{\bf Lemma 1.3} \em
A linear map $A_d:R\rt R_{n\times n}$ from the algebra $R$ into the
algebra of $n$ by $n$ matrices over $R$ given by the formula
$$
A_d(v)^i_k= D_k(vx^i)- D_k(v)x^i  \eqno  (3)
$$
is an algebra homomorphism i.e.
$$
A_d(uv)^i_k =A_d(u)^l_k A_d(v)^i_l  \eqno  (4)
$$
\em
{\it Proof:} Let $v \in R$. The left multiplication $A(v):\omega \mapsto
v\omega$ is an endomorphism of the right module $\Omega_d(R)$. Ring of all
endomorphisms of any free module of rank $n$ is isomorphic to the ring of
all $n$ by $n$ matrices. Therefore, we can find a homomorphism $A: R
\rt R_{n\times n}$ defined by the formulae
$$
v\,dx^i=dx^k\c A(v)_k^i
$$
By the Leibnitz rule we have
$$
v\,dx^i=d(vx^i) - d(v)x^i= dx^k[D_k(vx^i)-D_k(v)x^i]
$$
therefore,
$$
dx^k\c A_d(v)^i_k=dx^k\c A(v)^i_k
$$
i.e. $A_d=A$ and $A_d$ is also a homomorphism.\medskip \hfill $\Box$

Let us consider a linear map $D:R\rt R^n$ of $R$ to the space of columns
of height $n$ acted by the formula
$$
D(v)= \left( \ba{c}
D_1(v)\\ \vdots \\ D_n(v)
 \ea \right)  \ \ \ \!\!\!\! ,\;\; i.e. \ \ \ D= \left( \ba{c}
D_1\\ \vdots \\ D_n
\ea \right) \medskip
$$\\
{\bf Proposition 1.4} \em
The map $D$ and homomorphism $A_d$ are connected by the relation
$$
D(uv)=D(u)v + A_d(u)D(v)   \eqno    (5)
$$\\ \em
{\it Proof: } We have
$$
u\,dv=u\,dx^i\c D_i(v)=dx^k\c A_d(u)^i_k D_i(v)
$$
and
$$
dx^k\c D_k(uv)=D_k(u)v\,+\,A_d(u)^i_kD_i(v)
$$
i.e. by the uniqueness condition
$$
D_k(uv)=D_k(u)v\,+\,A_d(u)^i_kD_i(v)
$$ \hfill $\Box$

The inverse statement is also valid \medskip \\
{\bf Proposition 1.5} \em
Let $R$ be an algebra generated by elements $x^1,\ld ,x^n$
and $A:R \rt R_{n \times n}$ be an algebra homomorphism. If $D:R\rt R^n$
is a linear map such that
$$
D_k(x^i)=\delta^i_k  \eqno   (6)
$$
$$
D(uv)=D(u)v+A(u)D(v),	\eqno	(7)
$$
then the map $\Delta:v\mapsto dx^k\c D_k(v)$ is a free differential, where
$\Omega_\Delta (R)=\sum dx^i\c R$ is a free right module with the left module
structure defined by commutation rule, i.e. $A_\Delta =A$.\\ \em
{\it Proof: } We have to prove $\Delta(x^i)=dx^i$ and the Leibnitz formulae.
First equality follows from (6) and definition of $\Delta$. Finally
$$
\Delta(uv)=dx^k\c D(uv)=dx^k\c [D_k(u)v\,+\,A_d(u)^i_kD_i(v)]=
\Delta(u)v\,+\,u\Delta(v).
$$
\medskip \hfill $\Box$

A natural question concerning Proposition 1.5 arises here. If a homomorphism
$A$ is given, then formula (7) allows one to calculate partial derivatives
of a product in terms of its factors. That fact and formula (6) show that for
a given $A$ there exists not more then one $D$ satisfying formulas (6) and (7).
It is not
clear yet whether or not there exists at least one $D$ of such a type. Thus,
our
first
task is to describe these homomorphisms of $A$ for which there exist free
differentials with $A_d=A$. \medskip \\
{\bf Theorem1.6} \em
Let $R=F<x^1,\ld ,x^n>$ be a free algebra generated by
$x^1,\ld ,x^n$ and $A^1,\ld ,A^n$ be any set of $n\times n$ matrices over
$R$. There exists a unique free differential $d$ such that $A_d(x^i)=A^i$.\\
\em
{\it Proof: } The map $x^k \mapsto A^k$ can be uniquely extended to a
homomorphism of algebras $A:R \rt R_{n\times n}$. Let us define a map
$D$ on monomials in $x^1,\ld ,x^n$ by induction on its degree.  Let
$D_k(x^i)=\delta_k^i$ and
$$
D_k(x^iv)=\delta_k^i v\,+\,A(x^i)^j_kD_j(v),\ \ \ \ where \ A(x^i)=A^i
$$
We have to prove formulae (7) for arbitrary elements $u,\,v$. It can be done
by induction on degree of a monomial $u$.\\
If this degree is equal to one then (7) implies required result. Let
$u=x^iu_1$. Then by the equality $A(u)=A(x^iu_1)=A(x^i)A(u_1)$ and by
induction supposition we have
$$
D_k(uv)=D_k(x^iu_1v)=\delta^i_ku_1v\,+\,A(x^i)^j_kD_j(u_1v)=
$$
$$
=[\delta^i_ku_1\,+\,A(x^i)^j_kD_j(u_1)]v\,+\,A(x^i)^l_kA(u_1)^j_lD_j(v)=
D_k(u)v\,+\,A(u)^j_kD_j(v)
$$
\nopagebreak \hfill $\Box$

Let now $R$ be a non-free algebra defined by the set of generators $x^1,\ld,x^n$
and the set of relations $f_m(x^1,\ld ,x^n)=0,\ m\in M\ i.e.\ R=\R/ \I$, where
$\R=F<\x^1,\ld ,\x^n>$ is a free algebra and $\I$ is its ideal generated by
elements $f_m(\x^1,\ld ,\x^n),\,m\in M$.

Let us denote by $\pi$ the natural projection $\R \rt R$ such that
$\pi(\x^i)=x^i$. Since $R_{n\times n}= R\ot F_{n\times n}$, $\pi$
defines an epimorphism $\pih:\R_{n\times n}\rt R_{n\times n}$ by the formulae
$\pih=\pi \ot id$, where $id: F_{n\times n}\rt F_{n\times n}$ is the
identity map.\\ 

If $A:R\rt R_{n\times n}$ is any homomorphism of algebras, then we have the
following diagram of algebra homomorphism
$$
\ba{ccccc}
\R & \stackrel{\A}{\longrightarrow} & \R\ot F_{n\times n} & = & \R_{n\times
n}\\
\pi \downarrow &   & \downarrow \pi \ot id & = & \downarrow \pih \\
R & \stackrel{A}{\longrightarrow} & R\ot F_{n\times n} & = & F_{n\times n}
\ea  \eqno    (8)
$$

Let us choose for any generator $x^i$ an arbitrary element $\A^i\in \R$ such
that $\pih(\A^i)=A^i$ (recall that $\pih$ is epimorphism). Then the map
$\x^i\mapsto  \A^i$ can be extended to an algebra homomorphism $\A:\R
\rt \R_{n\times n}$ (recall that $\x^1,\ld ,\x^n$ are free variables). That
homomorphism completes (8) to a commutative diagram. For any relation $f_m
(\x^1,\ld ,\x^n)$ we have:
$$
\pih(\A(f_m(\x^1,\ld ,\x^n))) = A(\pi(f_m(\x^1,\ld ,\x^n)))=0. \eqno  (9)
$$
Furthermore,
$$
ker\pih=ker(\pi\ot id)=ker\pi\ot F_{n\times n}=I_{n\times n},
$$
and finally
$$
\A(f_m(\x^1,\ld ,\x^n)) \in ker\pih=I_{n\times n}.
$$
Theorem 1.6 claims that for the homomorphism $\A:\R\rt \R_{n\times n}$ there
exi
a unique free differential $\d$ of the free algebra $\R$. \medskip \\
{\bf Definition 1.7}
The differential $\d$ is called a {\it cover} differential with
respect to the homomorphism $A:R\rt R_{n\times n}$\,. \medskip \\
Thus we have proved: \smallskip \\
{\bf Theorem 1.8} \em
For any homomorphism $A:R\rt R_{n\times n}$ there there exists
a cover differential $\d$ of the algebra $\R$. \medskip \\ \em
{\bf Proposition 1.9} \em
An algebra $R$ with generators $x^1,\ld ,x^n$ and the set of
defining relations $\{f_m,\,m\in M \}$ has a free differential with respect to
a homomorphism $A:R\rt R_{n\times n}$ if and only if
$$
\hat{D}_k(f_m(\x^1,\ld ,\x^n))\in \I
$$
where $\hat{D}_k$ are partial derivatives of the cover differential $\d$.\\ \em
{\it Proof: } Let the free differential exist. We claim that the diagram
$$
\ba{ccc}
\R & \stackrel{\D}{\longrightarrow} & \R^n\\
\pi \downarrow &   & \downarrow \pi^n\\
R & \stackrel{D}{\longrightarrow} & R^n
\ea
$$
is commutative. Indeed, the difference $\Delta=D\circ \pi \,-\,\pi^n\circ \D$
acts trivially on generators:
$$
\Delta_k(\x^i)=D_k\pi(\x^i)\,-\,\pi \D_k(\x^i)=\delta^i_k\,-\,\pi(\delta^i_k)=0
$$
Commutativity of (8) implies $A(\pi(f))=\pih(A(f))$ and by (7) we have
$$
\Delta(fh)=D(\pi f\c \pi h)\,-\,\pi^n\D(fh)=
$$
$$
=D(\pi f)\pi h\,+\,A(\pi f)D(\pi h)\,-\,\pi^n(\D\c h\,+\,\A f\c \D h)=
$$
$$
=\Delta(f)\pi h\,+\,A(\pi f)\Delta(h)
$$
By evident induction, $\Delta=0$. \\
Finally for any relation $f_m$ we have
$$
\pi \D(f_m(\x^1,\ld ,\x^n))=D(\pi(f_m(\x^1,\ld ,\x^m)))=0
$$
i.e. $\D_k(f_m)\in ker \pi=\I$. \\
Inversely, if $\D_k(f_m)\in ker \pi=\I$ then we have
$$
\D(uf_mv)=\D(u)f_mv\,+\,\A(u)\D(f_m)v\,+\,\A(u)\A(f_m)\D(v)
$$
$$
\equiv \A(u)\A(f_m)\D(v)\ \ \ \ \ \ \ (mod\,I )
$$
By (8) one has $\pih \A(f_m)=A(\pi f_m)=0$ and $\A(f_m)\in ker\,\pih
=I_{n\times n}$. Therefore $\D_k:\R\rt \R$ induce maps $D_k:\R/\I
\rt \R  \stackrel{\pi}{\longrightarrow}R$ in such a way that
$D\circ \pi=\pi^n\circ \D$. Finally for arbitrary $u=\pi f\in R$ and
$v=\pi h\in R$ we have
$$
D(uv)=D(\pi f\c \pi h)=\pi^n\D(f)h\,+\,\pi^n\A(f)\D(h)=
$$
$$
=D(\pi f)\pi h\,+\,A(\pi f)D(\pi h)=D(u)v\,+\,A(u)v
$$
and by Proposition 1.5 the proposition is proved. \medskip \hfill $\Box$
{\bf Corollary 1.10} \em
Let an algebra $R$ be defined by generators $x^1,\ld ,x^n$
and the set of homogeneous relations  $\{f_m \}$ of the same degree. If $A:R
\rt R_{n\times n}$ acts linearly on generators $A(x^j)^i_k=\al^{i j}_{kl}
x^l$, then for the pair $(R, A)$ there exists a free differential iff
for all $m$
$$
\medskip \d f_m=0\ \ .  \eqno  (10)
$$
\\ \em
{\bf Definition 1.11}
An ideal $I$ of a free algebra $\R=F<\x^1,\ld ,\x^n>$ is said to be
{\it comparable} with a homomorphism $A:\R\rt \R_{n\times n}$ if the factor
ring $\R/I$ has a free differential satisfying the commutation rules
$$
x^jdx^i=dx^k\c A(x^j)^i_k\ \ .  \eqno  (11)
$$\\

If an ideal $I$ is $A$-comparable, then Lemma 1.3 defines a homomorphism
$A:r\mapsto A^i_k(r)$ from the factor algebra into the matrix algebra over it.
Thanks to  Proposition 1.8, it follows that $I$ is $A$-invariant and $A$-stable
in the sense of the following definition: \smallskip \\
{\bf Definition 1.12}
An ideal $J$ of the algebra $\R$ is said to be $A$-{\it invariant}
if $A^i_k(J)\subseteq J$, where $A:r\mapsto A^i_k(r)$ is a homomorphism. An
ideal $I$ is said to be $A$-{\it stable} if $D_k(I)\subseteq I$ for any
of partial derivatives $D_k$ defined by a differential $d$ corresponding to
$A$ (see Theorem 1.6). \medskip \\

For any homomorphism $A$ there exists the largest $A$-comparable ideal $I(A)$
-- the sum of all comparable ideals. It is is again $A$-comparable
because a sum of invariant ideals is invariant and a sum of of stable ideals
is stable one.

Now, we are going to describe the ideal $I(A)$ in the homogeneous case. If a
homomorphism $A$ preserves a degree, then it must act linearly on generators
$A^i_k(\x^j)=\al^{ij}_{kl}\x^l$. Therefore, the homomorphism $A$ is defined by
t
2-covariant 2-contravariant tensor $A=\al^{ij}_{kl}$. \medskip \\
{\bf Theorem 1.13} \em
For any 2-covariant 2-contravariant tensor $A=\al^{ij}_{kl}$
the ideal $I(A)$ can be constructed by induction as the homogeneous space
$I(A)=
I_1(A)+I_2(A)+I_3(A)+,\cdots$ in the following way:
\begin{enumerate}
   \item  $I_1(A)=0$
   \item  Assume that $I_{s-1}(A)$ has been defined and $U_s$ be a space of all
    polynomials $m$ of degree $s$ such that $D_k(m)\in I_{s-1}(A)$ for all
    $k.\ 1\leq k\leq n$. Then $I_s(A)$ is the largest $A$-invariant
    subspace of $U_s$.
\end{enumerate} \em
{\it Proof: } First of all, we should note that the maximal
$A$--comparable ideal has to be homogenous (graded). It is sufficient
to prove that every $A$--comparable ideal is contained in the homogenous
one. Since our free algebra $\R=\R_1+\R_2+\ld$ is graded, every element
$u \in \R$ has unique decomposition $u=u_1+u_2+\ld$ into homogenous
components. Let $J$ be an arbitrary $A$--comparable ideal. Define $J_s=
\{u_s: u \in J \}$. For $u\in J$ one has $A_k^i(u)=A_k^i(u_1)+A_k^i(u_2)
+\ld \,\in A_k^i(J)\subseteq J$ and $deg A_k^i(u_s)=deg u_s =s$. Therefore
$A_k^i(J_s)\subseteq J_s$. Analogously, $D_k(u)=D_k(u_1)+D_k(u_2)+\ld\,
\in D_k(J)\subseteq J$ and $deg D_k(u_s)=s-1$. So $D_k(J_s)\subseteq J_{s-1}$
and the sum $J_1+J_2+\ld$ is an $A$--comparable subset. Similarly,
$\R_tJ_s\R_p\subseteq J_{t+s+p}$, hence $J_1+J_2+\ld$ is an ideal in $\R$.\\
Next step is to prove that $I(A)$ is an ideal.
It is sufficient to show that $I_{s-1}\x^i+\x^jI_{s-1}\subseteq I_s \ \forall
\,i,\,j$. Let $V$be the space generated by the variables $\x^1,\ld ,\x^n$.
Let us prove by induction that $I_{s-1}V+VI_{s-1}\subseteq I_s$. We have
$$
D_k(I_{s-1}V+VI_{s-1})\subseteq D_k(I_{s-1})V+A^j_k(I_{s-1})D_j(V)+
D_k(V)I_{s-1}+
$$
$$
+A^j_k(V)D_j(I_{s-1})
\subseteq I_{s-2}\c V+I_{s-1}+I_{s-1}+V\c I_{s-2}\subset I_{s-1}
$$
It follows that $I_{s-1}V+VI_{s-1}\subseteq U_s$. Finally, the space
$I_{s-1}V+VI_{s-1}$ is $A^j_k-$ invariant as so are $I_{s-1}$ and $V$.
Therefore $I_{s-1}V+VI_{s-1}\subset I_s$ and $I$ is an ideal. \\
Let now $J=J_1+J_2 \ld$ be an arbitrary (graded) $A-$comparable ideal.
We are going to prove by induction that $J_s\subseteq I_s$. Let
$u=\beta_k\x^k\in J_1$. Then $\beta_kx^k=0$ in factor--ring $\R/J$.
Therefore $\beta_kdx^k=0$ and $\beta_k=0$ by uniqueness condition. So
$u=0$ in the free algebra and $0=J_1=I_1$.\\
Let $J_{s-1}\subseteq I_{s-1}$. By the Proposition 1.9 one has $D_k(J_s)
\subset J$. All elements from $D_k(J_s)$ have degree equal to $s-1$.
Therefore $D_k(J_s)\subseteq J_{s-1}\subseteq I_{s-1}$ and by the definition
of $U_s$ we have $J_s\subseteq U_s$. Finally $J_s$ is $A$--invariant
space and by the definition of the space $I_s$ we obtained $J_s\subseteq
I_s$. \medskip \hfill $\Box$\\

Let us denote by $\R_A$ the factor algebra $\R/I(A)$. In some sense $\R_A$
is an {\it optimal} algebra which has a free differential with respect to the
commutation rule $A$. Indeed, Theorem 1.13 shows in particular that if a
homogeneous element is such that all elements of the invariant subspace
generated by it have all partial derivatives equal to zero, then that element
vanishes in the optimal algebra.

We also have proved that there exists maximal algebra which has a free
differential with any given commutation rule (this is the free algebra,
see Theorem 1.6). Of course, it is very interesting to consider a number of
concrete commutation rules $A$ and related algebras $\R_A$. \medskip \\
{\bf Example 1.}
Let us consider the diagonal commutation rule: $x^jdx^i=dx^i\c
q^{ij}x^j$, with the symmetry condition $q^{ij}q^{ji}=1, i\neq j$. If none
of the coefficients $q^{ij}$ is a root of a polynomial of the type
$\la^{[m]}\doteq \la^{m-1}+\la^{m-2}+\cdots +1$, then the optimal algebra $\R
_A$ is equal to $F<x^1,\ld ,x^n>/\{q^{ij}x^ix^j=x^jx^i, \ i<j \}$.\\
If $(q^{ii})^{[m_i]}=0,\ 1\leq i\leq s$ with minimal $m_i$ then \\
$\R_A=F<x^1,\ld ,x^n>/\{q^{ij}x^ix^j=x^jx^i,\ i<j,\ (x^i)^{m_i}=0,\
1\leq i\leq s \}$. \medskip \\
{\bf Example 2.}
Let $A=0$ i.e. $x^idx^j=0$. Then $\d$ is a homomorphism of right
modules and the optimal algebra is free $\R_A=\R$.\medskip \\
{\bf Example 3.}
Let $x^idx^j=-dx^i\c x^j$. Then the optimal algebra is the
smallest possible algebra generated by the space $V$ i.e. $\R_A=F<x^1,\ld
,x^n>/ \{x^ix^j=0 \}$. \medskip \\
{\bf Example 4.}
Let $x^1dx^1= dx^1\c (\al_2x^2+\cdots +\al_nx^n)$ and
$x^idx^j=-dx^i\c x^j$ if $i\neq 1\ or \ j\neq 1$. Then the optimal algebra
is almost isomorphic to the ring of polynomials in one variable. More
precisely,
$\R_A=F<x^1,\ld ,x^n>/ \{x^ix^j=0, \ unless \ \,i=j=1 \}$. \medskip \\
{\bf Example 5.}
If $n=2$ and $x^1dx^1=dx^1\c \mu x^2, \ \ x^1dx^2=-dx^1\c x^2, \ \
x^2dx^1=-dx^2x^1, \ \ x^2dx^2=dx^2\c \la x^1$, then the optimal algebra is
isomorphic to the direct sum of two copies of the polynomial algebra
$\R_A=F<x^1, x^2>/ \{x^1x^2=x^2x^1=0 \}$. \medskip \\

Finally, we can formulate result which describes numbers of commutation
rules in two variables for which the optimal algebra is commutative.
\smallskip \\
{\bf Theorem 1.14} \em
In the two variable case, the following five series have commutative optimal
algebra:\\

\begin{enumerate}
 \item \ \ \
$
x^1dx^1=dx^1\c u\,+\,dx^2\c s ,\ \ \ \
x^1dx^2=dx^1\c w\,+\,dx^2\c (\la s+x^1),
$
$$
x^2dx^1=dx^1\c(w+x^2)\,+\,dx^2\c(\la s),
$$
$$
x^2dx^2=dx^1\c (\la w)\,+\,dx^2\c (\la^2s-\la u+w+\la x^1+x^2);
$$
  \item \ \ \
$
x^1dx^1=dx^1\c(x^1+\la w+s)\,+\,dx^2\c w, \ \ \ \
x^1dx^2=dx^1\c \ga w\,+\,dx^2\c (x^1+s),
$
$$
x^2dx^1=dx^1\c (x^2+\ga w)\,+\,dx^2\c s,\ \ \ \ \
x^2dx^2=dx^1\c \ga s\,+\,dx^2\c (x^2+\ga w-\la s);
$$
  \item \ \ \
$
x^1dx^1=dx^1\c (x^1+\ga w),\ \ \ \ \ x^1dx^2=dx^1\c w\,+\,dx^2\c x^1,
$
$$
x^2dx^1=dx^1\c (x^2+w),\ \ \ \ \ x^2dx^2=dx^1\c s\,+\,dx^2\c (x^2+w-\ga s);
$$
  \item \ \ \
$
x^1dx^1=dx^1\c u,\ \ \ \ \ \ x^1dx^2=dx^2\c x^1,
$
$$
x^2dx^1=dx^1\c x^2,\ \ \ \ \ \ x^2dx^2=dx^2\c v;
$$
  \item\ \ \
$
x^1dx^1=dx^1\c u,\ \ \ \ \ \  x^1dx^2=dx^2\c u,
$
$$
x^2dx^1=dx^1\c x^2\,+\,dx^2\c (u-x^1),\ \ \ \ \ x^2dx^2=dx^1\c w\,+\,dx^2\c v.
$$
\end{enumerate}
where $u, v, w, s$ are arbitrary elements from the space $V$ and $\la, \ga$
are parameters from the base field.\smallskip \\ \em
{\bf Remark. }The above series are not independent. For example, the standard
Newton--Leibnitz calculi ($x^idx^j=dx^j\c x^i$) belongs, as a special case
to each of them, by putting $s=w=0, u=x^1, v=x^2$. More detailed discussion
of the above examples and classification theorem for calculi with a commutative
optimal algebra will be given elsewhere \cite{next}.


\begin{thebibliography}{}
\bibitem{Wess}
J. Wess and B. Zumino, {\it Covariant differential calculus on the quantum
hyperplane}. Nuclear Physics {\bf 18B} (1990), p.303 .
\bibitem{next}
A. Borowiec, V. K. Kharchenko and Z. Oziewicz {\it Differentials
with Uniqueness Property} -- in preparation.
\end{thebibliography}
\end{document}